\documentclass[conference,a4paper]{IEEEtran}

\IEEEoverridecommandlockouts

\usepackage[english]{babel}
\usepackage{amsmath, amssymb, amsbsy}
\usepackage{cite}
\usepackage{enumitem}
\usepackage{colortbl}
\usepackage{amsthm}

\makeatletter
\newcommand\footnoteref[1]{\protected@xdef\@thefnmark{\ref{#1}}\@footnotemark}
\makeatother

\newtheorem{theorem}{Theorem}
\newtheorem{lemma}[theorem]{Lemma}

\newtheorem{proposition}[theorem]{Proposition}

\newtheorem{remark}[theorem]{Remark}
\newtheorem{openproblem}{Open Research Problem}
\newtheorem{definition}{Definition}

\newenvironment{mymatrix}{\begin{bmatrix}} {\end{bmatrix} }

\def\ve#1{{\mathchoice{\mbox{\boldmath$\displaystyle #1$}}%
              {\mbox{\boldmath$\textstyle #1$}}%
              {\mbox{\boldmath$\scriptstyle #1$}}%
              {\mbox{\boldmath$\scriptscriptstyle #1$}}}}

\definecolor{mygreen}{rgb}{0,0.6,0}

\renewcommand{\vec}[1]{\ensuremath{\boldsymbol{#1}}}

\newcommand{\MoormatExplicit}[3]{
	\begin{mymatrix}
		#1_{1} & #1_{2} & \dots& #1_{#3}\\
		#1_{1}^{[1]} & #1_{2}^{[1]} & \dots& #1_{#3}^{[1]}\\
		\vdots &\vdots&\ddots& \vdots\\
		#1_{1}^{[#2-1]} & #1_{2}^{[#2-1]} & \dots& #1_{#3}^{[#2-1]}\\
	\end{mymatrix}}

\newcommand{\Fq}{\ensuremath{\mathbb{F}_q}}
\newcommand{\Fqm}{\ensuremath{\mathbb{F}_{q^m}}}
\newcommand{\Fqml}{\ensuremath{\mathbb{F}_{q^{m\ell}}}}

\DeclareMathOperator{\extsmallfield}{ext}
\DeclareMathOperator{\rank}{rk}

\newcommand{\LambdaOp}{\Lambda}
\newcommand{\Code}{\mathcal{C}}
\newcommand{\qpow}[1]{^{[#1]}}
\newcommand{\rspace}[1]{\mathcal{R}\begin{pmatrix}#1\end{pmatrix}}

\newcommand{\cipher}{\ve{Y}}
\newcommand{\msg}{\ve{M}}

\newcommand{\tpub}{\ensuremath{t_{\mathsf{pub}}}}

\newcommand{\tprime}{\ensuremath{t^{\prime}}}
\newcommand{\mycode}[1]{\ensuremath{\mathcal{#1}}}

\newcommand{\Gabcode}[2]{\ensuremath{\mathcal{G}(#1,#2)}}
\newcommand{\IntGabcode}[1]{\ensuremath{\mathcal{IG}(#1)}}

\renewcommand{\S}{\ve{S}}
\renewcommand{\H}{\ve{H}}
\newcommand{\y}{\ve{y}}
\renewcommand{\a}{\ve{a}}
\renewcommand{\b}{\ve{b}}
\newcommand{\e}{\ve{e}}
\renewcommand{\v}{\ve{v}}
\newcommand{\g}{\ve{g}}

\renewcommand{\c}{\ve{c}}
\renewcommand{\e}{\ve{e}}

\newcommand{\G}{\ve{G}}
\renewcommand{\P}{\ve{P}}

\newcommand{\M}{\ve{M}}
\newcommand{\U}{\ve{U}}
\newcommand{\E}{\ve{E}}
\newcommand{\HE}{\ve{H}_{\text{E}}}
\newcommand{\Gpub}{\ve{G}_\mathrm{pub}}
\newcommand{\Hpub}{\ve{H}_\mathrm{pub}}

\newcommand{\A}{\ve{A}}
\newcommand{\B}{\ve{B}}

\newcommand{\dE}{d_\mathrm{E}}
\renewcommand{\AE}{\A}
\newcommand{\BE}{\B}
\newcommand{\dR}{\mathrm{d}_\mathrm{R}}

\newcommand{\Aprime}{\A^{\prime}}
\newcommand{\Bprime}{\B^{\prime}}

\newcommand{\Etilde}{\tilde{\E}}
\newcommand{\Btilde}{\tilde{\B}}

\newcommand{\Pfail}{P_{\text{f}}}

\newcommand{\Binv}{\B_\mathrm{inv}}
\newcommand{\Bker}{\B_\mathrm{ker}}
\newcommand{\GaugY}{\G_\mathrm{aug}^{\cipher}}
\newcommand{\GaugE}{\G_\mathrm{aug}^{\E}}
\newcommand{\CodeE}{\Code_\E}
\newcommand{\Caug}{\Code_\mathrm{aug}}

\newcommand{\x}{\ve{x}}
\newcommand{\X}{\ve{X}}
\newcommand{\Y}{\ve{Y}}
\newcommand{\I}{\ve{I}}
\renewcommand{\a}{\ve{a}}

\begin{document}

\title{Interleaving Loidreau's Rank-Metric Cryptosystem\\  
  \thanks{
    This work was supported by the European Research Council (ERC) under the European Union’s Horizon 2020 research and innovation programme (grant agreement No~801434), and by the German Israeli Project Cooperation (DIP) grant no.~KR3517/9-1.
  }
}

\author{\IEEEauthorblockN{Julian Renner, Sven Puchinger, Antonia Wachter-Zeh}
\IEEEauthorblockA{\textit{Institute for Communications Engineering} \\
\textit{Technical University of Munich (TUM)}\\
Munich, Germany \\
\{julian.renner,sven.puchinger,antonia.wachter-zeh\}@tum.de}
}

\maketitle

\begin{abstract}
We propose and analyze an interleaved variant of Loidreau's rank-metric cryptosystem based on rank multipliers.
We analyze and adapt several attacks on the system, propose design rules, and study weak keys.
Finding secure instances requires near-MRD rank-metric codes which are not investigated in the literature. Thus, we propose a random code construction that makes use of the fact that short random codes over large fields are MRD with high probability.
We derive an upper bound on the decryption failure rate and give example parameters for potential key size reduction.
\end{abstract}


\section{Introduction}
Code-based cryptosystems have gained large attention in the last years since they are potentially resistant to quantum computer attacks, in contrast to currently-used number theoretic systems like RSA or ElGamal.
The most famous code-based cryptosystem is the one by McEliece \cite{mceliece1978public}, which is based on the hardness of decoding in a generic code.

Recently, \cite{elleuch2018interleaved} introduced a system which can potentially reduce the key size of the original McEliece cryptosystem. The proposed system uses the same public key as the original system, but changes the cipher to a corrupted codeword of an interleaved code.
Hence, key attacks are as hard as on the original McEliece system and one potentially obtains a better resistance against generic decoding since the interleaved code can correct significantly more errors than a single Goppa code.
However, Tillich \cite{tillich2018attack} found an attack, which is more efficient than generic decoding if the error is not chosen carefully.
A repair against Tillich's attack was proposed in \cite{holzbaur2019on}.

Rank-metric codes are a promising candidate for code-based cryptography since generic decoding in the rank metric appears to be much harder than generic decoding in the Hamming metric.
Hence, they provide significantly smaller key sizes at the same level of security against generic decoding.
The rank metric was first considered in a McEliece-like scheme in \cite{gabidulin1991ideals} (\emph{Gabidulin--Paramonov--Tretjakov (GPT) cryptosystem}). There are several modifications of the GPT system \cite{gabidulin2001modified,gabidulin2003reducible,loidreau2010designing,rashwan2011security,gabidulin2008attacks,gabidulin2009improving,rashwan2010smart,loidreau2016evolution,Loidreau2017-NewRankMetricBased}, which are all based on hiding the structure of a Gabidulin code, the most famous family of rank-metric codes, from an attacker.
However, most of these systems are broken by Gibson's \cite{gibson1995severely} and Overbeck's \cite{overbeck2008structural} attacks, as well as modifications thereof.

The only Gabidulin-code-based GPT variant that has not been broken so far is the one by Loidreau \cite{Loidreau2017-NewRankMetricBased}.
There are also GPT variants based on other code classes, e.g., \cite{gaborit2013low,berger2017gabidulin}, as well as other types of rank-metric-code-based cryptosystems, e.g., \cite{faure2006new,wachter2018repairing,aguilar2018efficient}, which we will not consider here.

In this paper, we combine the ideas of the interleaved system in \cite{elleuch2018interleaved} with Loidreau's GPT variant \cite{Loidreau2017-NewRankMetricBased}.
We show that in principle, Loidreau's system can be interleaved using classical decoders for interleaved Gabidulin codes.
We also analyze the security of the new system, including an adaption of Tillich's attack to the rank metric.
Similar to \cite{holzbaur2019on}, we describe how Tillich's attack can be prevented by choosing the error matrix in a suitable way.
It turns out that the construction of (in this sense) secure errors requires rank-metric codes whose minimum distances are close to the Singleton bound.
We show that Gabidulin codes yield potentially insecure error patterns since the resulting error matrix can be distinguished from a random one. We further show that depending on the parameters, one can draw the error matrix in a random way and fulfill the requirements with high probability. For this choice of the error, we derive upper bounds on the decryption failure and present secure parameter sets that demonstrate the potential key size reduction.

\section{Preliminaries}
\subsection{Notations}
Let $q$ be a power of a prime and let
$\Fq$ denote the finite field of order $q$ and $\Fqm$ its extension field of order $q^m$. 
We use $\Fq^{m \times n}$ to denote the set of all $m\times n$ matrices over $\Fq$ and $\Fqm^n =\Fqm^{1 \times n}$ for the set of all row vectors of length $n$ over $\Fqm$. Rows and columns of $m\times n$-matrices are indexed by $1,\dots, m$ and $1,\dots, n$, where $A_{i,j}$ is the element in the $i$-th row and $j$-th column of the matrix $\A$ . Denote the set of integers $[a,b] = \{i: a \leq i \leq b\}$. By $\rank_q(\A)$ and $\rank_{q^m}(\A)$, we denote the rank of a matrix $\A$ over $\Fq$, respectively $\Fqm$.

For any $i$, we denote the $q$-power by $[i]:=q^i$.

Let $ \ve{\gamma} = \begin{mymatrix}\gamma_1,\gamma_2,\dots,\gamma_{m}\end{mymatrix}$ be an ordered basis of $\Fqm$ over $\Fq$. By utilizing the vector space isomorphism $\Fqm \cong \Fq^m$, we can relate each vector $\a \in \Fqm^n$ to a matrix $\A \in \Fq^{m \times n}$ according to
$\extsmallfield_{\boldsymbol{\gamma}}:\Fqm^{n} \rightarrow \Fq^{m \times n}\label{eq:mapping_smallfield},~\a = \begin{mymatrix}a_1,\hdots,a_n\end{mymatrix} \mapsto \A$,
where $a_j = \sum_{i=1}^{m} A_{i,j} \gamma_i, \quad \forall j \in [1,n]$.
Further, we extend the definition of $\extsmallfield_{\boldsymbol{\gamma}}$ to matrices by extending each row and then vertically concatenating the resulting matrices.

For a field $\mathbb{F}$, the vector space that is spanned by $\v_1,\hdots,\v_l \in \mathbb{F}^n$ is denoted by
$\langle\v_1,\hdots,\v_l\rangle_{\mathbb{F}} =\{\sum_{i=1}^{l}a_i\v_i \, : \, \ a_i \in \mathbb{F} \}$.  

The vector space that is spanned by the rows of the matrix $\A \in \mathbb{F}^{m\times n}$ is denoted by $\rspace{\A}$, i.e.,
$\rspace{\A} = \langle \begin{mymatrix} A_{1,1}, \hdots, A_{1,n} \end{mymatrix},\hdots,\begin{mymatrix} A_{m,1}, \hdots, A_{1m,n} \end{mymatrix} \rangle_{\mathbb{F}}$ .

The set of all $n \times n$ matrices which have only entries from $\mathcal{V}$ is denoted by $M_n(\mathcal{V})$, i.e.,
$M_n(\mathcal{V}) = \{ \A \in \Fqm^{n \times n}: A_{i,j} \in \mathcal{V} \}$.

The product space of the subspaces $\mathcal{A}$ and $\mathcal{B}$ is denoted by $\mathcal{A} \times \mathcal{B}$.

\subsection{Rank-Metric, Gabidulin and Interleaved Gabidulin Codes}
The \textit{rank norm} $\rank_q(\a)$ is the rank of the matrix representation $\A \in \Fq^{m \times n}$ over $\mathbb{F}_{q}$. 
The rank distance between $\a$ and $\b$ is
$\dR(\a,\b):= \rank_q(\a-\b) = \rank_q(\A-\B)$.
%
An $[n,k,d]$ linear code $\mycode{C}$ over $\Fqm$ is a $k$-dimensional subspace of $\Fqm^n$ and minimum rank distance $d$, i.e,
$d := \min_{\a \in \mycode{C} \setminus \{0\}} \lbrace \rank_q(\a) \rbrace$. 

Gabidulin codes \cite{Delsarte_1978,Gabidulin_TheoryOfCodes_1985,Roth_RankCodes_1991} are a class of rank-metric codes.
\begin{definition}[Gabidulin Code]
	A Gabidulin code $\Gabcode{n}{k}$ over $\Fqm$ of length $n \leq m$ 
	and dimension $k$ is defined by its $k \times n$ generator matrix
	\begin{equation*}
	\G = \MoormatExplicit{g}{k}{n},
	\end{equation*}
	where $\g=[g_1,g_2, \dots, g_{n}] \in \Fqm^n$, $\rank_q(\g) = n$. 
\end{definition}
In~\cite{Gabidulin_TheoryOfCodes_1985}, it is shown that Gabidulin codes are MRD codes, i.e., $d=n-k+1$, and can be decoded uniquely up to $t\leq\lfloor\frac{d-1}{2}\rfloor$.

Interleaved Gabidulin codes are a code class for which efficient decoders are known that are able to correct
$t \leq \lfloor \frac{\ell}{\ell+1}(n-k)\rfloor$ errors\footnote{In this setting, an error of weight $t$ is a $\ell \times n$ matrix over $\Fqm$ of $\Fq$-rank $t$. Note that this means that the tall $(\ell m) \times n$-matrix obtained by expanding the matrix component-wise over $\Fq$ has rank $t$.} with high probability, cf.~\cite{Loidreau_Overbeck_Interleaved_2006,Sidorenko2011SkewFeedback,WachterzehZeh-ListUniqueErrorErasureInterpolationInterleavedGabidulin_DCC2014}.

\begin{definition} [Interleaved Gabidulin Codes~\cite{Loidreau_Overbeck_Interleaved_2006}]
An interleaved Gabidulin code $\IntGabcode{\ell;n,k}$ over $\Fqm$ of length $n \leq m$, dimension $k \leq n$, and interleaving order $\ell$ is defined by
\begin{equation*}
\IntGabcode{\ell;n,k} \! := \!
\left\lbrace \!
\begin{mymatrix}
\vec{c}_{\mycode{G},1}^{\top} \hdots
\vec{c}_{\mycode{G},\ell}^{\top}
\end{mymatrix}^{\top}
\! : \! \vec{c}_{\mycode{G},i} \! \in \! \Gabcode{n}{k} ,\! \forall  i \! \in \! [1,\ell]
\right\rbrace.
\end{equation*}
\end{definition}

\subsection{Difficult Problems in Rank Metric}
In this section, we state difficult variants of the rank syndrome decoding (RSD) problem which can used for cryptography.

\begin{definition}[RSD Distribution]
Input: $q,n,k,w,m$\\
Choose uniformly at random
\begin{itemize}
\item $\H \xleftarrow{\$} \{\A \in \Fqm^{(n-k)\times n}: \rank_{q^m}(\A) = n-k \}$
\item $\x \xleftarrow{\$} \{\a \in \Fqm^{n} : \rank_{q}(\a) = w\} $
\end{itemize}
Output: $(\H,\H\x^{\top})$
  \end{definition}

\begin{definition}[Search RSD Problem]
Input: $(\H,\y^{\top})$ from the RSD Distribution\\
Goal: Find $\x \in \{\a \in \Fqm^{n} : \rank_{q}(\a) = w\}$ such that $\H\x^{\top} = \y^{\top}$
  \end{definition}
Note that the Syndrome Decoding Problem in Hamming Metric can be probabilistically reduced to Search RSD problem~\cite{gaborit2016Hardness}.

\begin{definition}[Interleaved RSD Distribution]
Input: $q,n,k,w,m,\ell$\\
Choose uniformly at random
\begin{itemize}
\item $\H \xleftarrow{\$} \{\A \in \Fqm^{(n-k)\times n}: \rank_{q^m}(\A) = n-k \}$
\item $\X \xleftarrow{\$} \{\B \in \Fqm^{\ell \times n} : \rank_{q}(\B) = w\} $
\end{itemize}
Output: $(\H,\H\X^{\top})$
\end{definition}

\begin{definition}[Interleaved Search RSD Problem]
Input: $(\H,\Y^{\top})$ from the Interleaved RSD Distribution\\
Goal: Find $\X \in \{\B \in \Fqm^{n} : \rank_{q}(\B) = w\}$ such that $\H\X^{\top} = \Y^{\top}$
\end{definition}
Note that the Interleaved Search RSD problem is similar to the problem proposed in~\cite[Definition 7]{gaborit2017IBE}. The only difference is that the rows of the matrix $\X$ in Interleaved RSD Distribution have the same row space whereas the rows of $\U^{\top}$ in~\cite[Definition 7]{gaborit2017IBE} have the same column space. For a small interleaving order $\ell$, the currently most efficient algorithm to solve both, the Interleaved Search RSD Problem and the problem given in~\cite[Definition 7]{gaborit2017IBE}, was presented in~\cite{tillich2018attack} and will be analyzed in Section~\ref{sec:attacks}. For a  high interleaving order $\ell\geq w$, the algorithm proposed in~\cite{puchinger2019DecHighInterleaved} is able to solve the Interleaved Search RSD Problem with high probability in polynomial time. For an interleaving order greater than $wk$, the algorithm proposed in~\cite{debris2019IBEAttack} is able to efficiently solve~\cite[Definition 7]{gaborit2017IBE}, see~\cite[Section 6.5]{debris2019IBEAttack}.

\section{Interleaving Loidreau's Cryptosystem}\label{sec:system}
The system that we propose is a McEliece-type system based on interleaving the rank-metric codes introduced in~\cite{Loidreau2017-NewRankMetricBased}.

To prove that decryption of the proposed system is successful with high probability, we need the following lemma.
\begin{lemma}\label{lemma:rank_multiplier}
  Let $\vec{P}\in M_n(\mathcal{V})$ be an invertible matrix with entries in a $\lambda$-dimensional $\Fq$-linear subspace $\mathcal{V}$ of $\Fqm$. Then
  \begin{equation*}
    \forall \vec{E}\in \Fqm^{\ell \times n}: \rank_q(\vec{E}\vec{P}) \leq \lambda \rank_q(\vec{E}).
    \end{equation*}
  \end{lemma}
  \begin{IEEEproof}
The proof is similar to \cite{Loidreau2017-NewRankMetricBased}.
Let $\ve{\gamma}^{\prime} = \begin{mymatrix}\gamma_1^{\prime},\dots,\gamma_{\ell}^{\prime}\end{mymatrix}$ be an ordered basis of $\Fqml$ over $\Fqm$, $\vec{e} =  \begin{mymatrix}e_1,\hdots, e_n \end{mymatrix} := \extsmallfield_{\boldsymbol{\gamma}^{\prime}}^{-1}(\E) \in \Fqml^n$ be of rank weight $t$, and $\langle e_1,\hdots, e_n \rangle_{\Fq} = \langle \epsilon_1,\hdots, \epsilon_t \rangle_{\Fq}$. Further, let $\nu_1,\hdots,\nu_{\lambda}$ be a basis of $\mathcal{V}$. The entries of the vector $\vec{e}\vec{P}$ belong to the vector space $\langle \epsilon_1\nu_1,\epsilon_2\nu_1,\hdots,\epsilon_t\nu_1,\epsilon_1\nu_2,\hdots, \epsilon_t\nu_{\lambda} \rangle_{\Fq}$ of dimension $\leq \lambda t$.
    \end{IEEEproof}

The system parameters are shown in Table~\ref{tab:parameters}. The key generation, encryption and decryption algorithms are as follows.

\begin{table}
  \caption{Summary of the Parameters}
  \vspace{-0.4cm}
\renewcommand{\arraystretch}{1.0} 
\begin{center}
\begin{tabular}{c|l|l}
Name & Use & Restriction \\
\hline
$q$ & small field size & prime power \\
$m$ & extension degree & $1 \leq m$ \\
$n$ & code length & $n \leq m$ \\
$k$ & code dimension & $k < n$ \\
$\lambda$ & dimension of the $\mathcal{V}$ & $\frac{n}{n-k}<\lambda \leq \lfloor \frac{n-k}{2} \rfloor$ \\
$\ell$ & interleaving order & $1 \leq \ell < \tpub$\\
$\tpub$ & error weight in ciphertext & $\tpub = \lfloor \frac{\ell}{\ell+1} \frac{n-k}{\lambda} \rfloor$
\end{tabular}
\end{center}
\label{tab:parameters}
\vspace{-0.8cm}
\end{table}

\subsection{Key Generation}
The keys are the same as in~\cite{Loidreau2017-NewRankMetricBased}, i.e.,
\begin{itemize}
\item $\G \in \Fqm^{k\times n}$ a generator matrix of a random $\Gabcode{n}{k}$,
\item $\S \in \Fqm^{k\times k}$, which is random and nonsingular
  \item $\P \in M_n(\mathcal{V}) \subset \Fqm^{n \times n}$, random and non-singular, where $\mathcal{V}$ is a random $\lambda$-dimensional $\Fq$-linear subspace of $\Fqm$.
  \end{itemize}
The public key is given by $\Gpub := \S \G \P^{-1}$.

\subsection{Encryption}

{\setlist[enumerate]{leftmargin=5mm}
\begin{enumerate}
\item Choose the error matrix $\E = \begin{mymatrix}  \e_1^{\top},\hdots, \e_\ell^{\top}  \end{mymatrix}^{\top}$
    randomly s.t.
\begin{equation}
  \rank_{q}(\E) = \Big\lfloor \frac{\ell}{\lambda(\ell+1)} (n-k) \Big\rfloor =: \tpub \ . \label{eq:decoding_condition}
\end{equation}

\item Compute the cipher
$  \cipher = \msg \Gpub + \E \in \Fqm^{l \times n}$,
where $\msg \in \Fqm^{\ell\times k }$ is the message matrix.  
  \end{enumerate}
}

\subsection{Decryption}

{\setlist[enumerate]{leftmargin=5mm}
\begin{enumerate}
\item Compute $\cipher \P = \msg   \S \G + \E^{\prime}$, where $\E^{\prime}:=\E \P $ and $\rank_q(\E^\prime)\leq \lfloor \frac{\ell}{\ell+1} (n-k) \rfloor$, cf. Lemma~\ref{lemma:rank_multiplier}.
\item Decode $\cipher \P$ in $\IntGabcode{\ell;n,k}$ to obtain $\M \S$.
\item Compute $\M \S \S^{-1} = \M$ to retrieve the message.
  \end{enumerate}
}

Assuming $\Gpub$ cannot be distinguished from a random matrix\footnote{The only known distinguisher~\cite{coggia2019AttackLoidreau} cannot be applied for a parameter choice according to Table~\ref{tab:parameters}.}, an attacker needs to generically decode the cipher to obtain the plain text. This is equal to solving the Interleaved Search RSD Problem.

\section{Attacks on the Cryptosystem}\label{sec:attacks}
We recall, analyze, and adapt known attacks on the systems in \cite{Loidreau2017-NewRankMetricBased,elleuch2018interleaved}.
Since the keys are the same as in~\cite{Loidreau2017-NewRankMetricBased}, key attacks are as hard as on the system in~\cite{Loidreau2017-NewRankMetricBased}.

\subsubsection{\textbf{(key attack)}} In \cite{Loidreau2017-NewRankMetricBased}, a structural attack is described, which is based on brute-forcing a number of $(\lambda-1)$-dimensional subspaces of $\Fq^m$. The work factor is given by\footnote{\label{note1} We divide the exponent by $2$ to obtain an estimate of the post-quantum work factor (presuming that Grover's algorithm can be applied).}
\begin{equation}
\text{WF}_{\text{Loi}} = q^{\frac{1}{2}((\lambda-1)m-(\lambda-1)^2)}.
\end{equation}

\subsubsection{\textbf{(decoding attack)}} The work factors\footnoteref{note1} of the algorithms that correct errors of rank $t$ in an arbitrary $[n,k]$ linear rank distance code over $\Fqm$ are denoted by $\text{WF}_{\text{Cha}} (t)$~\cite{chabaud1996rsd}, $\text{WF}_{\text{Our}}$~\cite{ourivski2002rsd}, $\text{WF}_{\text{Gab}}(t)$~\cite{gaborit2016rsd} and $\text{WF}_{\text{Ara}}(t)$~\cite{aragon:ISIT18}.

Tillich~\cite{tillich2018attack} proposed an attack on the interleaved Goppa codes system in \cite{elleuch2018interleaved}, which can be similarly applied here.
The augmented matrix of the public key and the cipher $\GaugY := \begin{mymatrix}\Gpub^{\top} \cipher^{\top} \end{mymatrix}^{\top}$
has the same row space as the matrix $\GaugE := \begin{mymatrix} \Gpub^{\top} \E^{\top} \end{mymatrix}^{\top} $. Thus, the row space $\Caug := \rspace{\GaugY}$ contains codewords of weight $\geq \dE$, where $\dE$ is the minimum rank distance of an error code spanned by the rows of $\E$, i.e.,
$\CodeE[n,\ell,\dE] := \rspace{\E}$.
Due to the restriction on the error matrix $\E$ in \eqref{eq:decoding_condition}, finding some non-zero element of the error code can, at least partially, recover the row space of the extended error matrix $\extsmallfield_{\ve{\gamma}}(\E)$ since
$\rspace{\extsmallfield_{\ve{\gamma}}(\E)} = \rspace{\extsmallfield_{\ve{\gamma}}(\e_1)} + \dots + \rspace{\extsmallfield_{\ve{\gamma}}(\e_\ell)}$.
The problem of finding low-rank-weight words was studied in \cite{hauteville2015new}, and is in principle equivalent to rank syndrome decoding.
In particular, it has a similar complexity if the weight of the low-weight words is as large as the error in rank syndrome decoding, i.e., the smallest-known work factor is
\begin{equation*}
  \text{WF}_{\E}\!=\!\min\{ \text{WF}_{\text{Cha}}(\dE),\!\text{WF}_{\text{Our}}(\dE),\!   \text{WF}_{\text{Gab}}(\dE),\!\text{WF}_{\text{Ara}}(\dE) \}.
\end{equation*}

Note that since each row of $\cipher$ is a codeword corrupted by an error of rank at least $\dE$, the row-wise rank syndrome decoding has a complexity of at least $\text{WF}_{\E}$. Further, this attack has a higher complexity than generic decoding in Loidreau's original system with the same public key if and only if $\dE>\tfrac{d-1}{2\lambda}$.

\subsubsection{\textbf{(decoding attack)}} In~\cite{puchinger2019DecHighInterleaved}, a polynomial-time decoding algorithm is proposed that works for arbitrary interleaved codes of interleaving degree $\ell\geq \tpub$ and error matrices of full rank. However in case of $\ell<\tpub$, one must brute-force through the solution space of a linear system of equations, whose size is exponential in $m(\tpub-\ell)$. By choosing the parameters according to Table~\ref{tab:parameters}, this attack is not efficient.

\begin{table*}[t!]
  \caption{Comparison of Loidreau's system with the presented interleaved codes system.}
  \vspace{-0.25cm}
  \setlength{\tabcolsep}{7pt}
\renewcommand{\arraystretch}{0.9}
	\begin{center}
		\begin{tabular}{l|c|c|c|c|c|c|c||c|c|c||c|c|r }
                  Method &$q$ & $k$ & $n$ & $m$ & $\lambda$ & $\ell$ & $\tpub$ & $\text{WF}_{\text{Loi}}$ & $\text{WF}_{\E}$ & $\text{WF}_{\AE}$ & Rate & $\Pfail$ & Key size  \\
                  \hline \hline
Classic & $16$ & $11$ & $27$ & $42$ & $2$ & $1$ & $4$ & $82.00$ & $80.38$ & $\infty$ & $0.41$ & $-\infty$ & $3.70$ KB\\
Interleaved & $16$ & $9$ & $27$ & $42$ & $2$ & $2$ & $6$ & $82.00$ & $86.48$ & $119.00$ & $0.33$ & $-166.00$ & $3.40$ KB\\
            \hline      
                  Classic & $16$ & $14$ & $34$ & $66$ & $2$ & $1$ & $5$ & $130.00$ & $128.39$ & $\infty$ & $0.41$ & $-\infty$ & $9.24$ KB\\
                  Interleaved & $16$ & $13$ & $31$ & $66$ & $2$ & $2$ & $6$ & $130.00$  & $128.07$ & $215.00$ & $0.42$ & $-266.00$ & $7.72$ KB\\
                  \hline
                  Classic  & $16$ & $23$ & $53$ & $62$ & $3$ & $1$ & $5$ & $240.00$  & $198.58$ & $\infty$ & $0.43$ & $-\infty$ & $21.39$ KB\\
                  Interleaved & $16$ & $22$ & $49$ & $62$ & $3$ & $2$ & $6$ & $240.00$ &  $200.34$ & $199.00$ & $0.45$ & $-246.00$ & $18.41$ KB\\
                  \hline

                  Classic  & $16$ & $30$ & $60$ & $68$ & $3$ & $1$ & $5$ & $264.00$ & $256.98$ & $\infty$ & $0.50$ & $-\infty$ & $30.60$ KB\\  
                  Interleaved & $16$ & $28$ & $55$ & $77$ & $3$ & $2$ & $6$ & $300.00$ & $257.77$ & $259.00$ & $0.51$ & $-306.00 $ & $29.11$ KB\\
		\end{tabular}
	\end{center}
	\label{tab:security_para}
	\vspace{-0.7cm}
\end{table*}

\section{Construction of the Error Matrix}

We have seen in the previous section that in order to resist Tillich's attack, the rows of the error matrix $\E$ must span a code of large minimum rank distance, i.e., $\E$ must be a generator matrix of an $[n,\ell,\dE>\frac{d-1}{2}]$ code.
The following statement shows how to construct such a code that still fulfills the decoding condition \eqref{eq:decoding_condition}, which is necessary for successful decryption.

\begin{theorem}\label{thm:E_construction}
Let the error matrix be given by
\begin{equation*}
\E
=
\AE \cdot \BE \ \in\Fqm^{\ell \times n} ,
\end{equation*}
where $\AE \in \Fqm^{\ell \times \tpub}$ is a generator matrix of a $[\tpub,\ell,\dE]$ code and has full $\Fq$-rank and $\BE \in \Fq^{\tpub \times n}$ has full rank.
Then, $\E$ fulfills \eqref{eq:decoding_condition} and is a generator matrix of an $[n,\ell,\dE]$ code. Also, $\E$ and any row of $\E$ has $\Fq$-rank at least~$\dE$.
\end{theorem}

\begin{IEEEproof}
Since $\AE$ has $\tpub$ columns, its $\Fq$-rank is at most $\tpub$.
Multiplication by the full-rank $\Fq$-matrix $\BE$ from the right does not change the $\Fq$-rank, so $\rank_q(\E)\leq \tpub$ and \eqref{eq:decoding_condition} is satisfied.

To prove that the error matrix spans an $[n,\ell,\dE]$ code, we first observe that the length of vectors in the row space of $\E$ is $n$ and its $\Fqm$-rank is $\ell$ (since $\AE$ has full $\Fqm$-rank and multiplication by the full-rank matrix $\BE$ does not change this rank). Thus it is a code of length $n$ and dimension $\ell$ over~$\Fqm$.

As for the minimum distance, we have the following.
Let $\c_1,\c_2$ be two distinct vectors in the row space of $\E$. Then, we can write them as $\c_i = \a_i \cdot \BE$, where $\a_1,\a_2$ are in the row space of $\AE$. Since the $\c_i$ are distinct, so are the $\a_i$. Furthermore, we have $\dR(\a_1,\a_2) \geq \dE$. Since $\BE$ is a full-rank matrix over $\Fq$, multiplication by it does not change the rank of a word. Hence,
$\dR(\c_1,\c_2) = \dR(\a_1 \BE,\a_2 \BE) = \dR(\a_1,\a_2) \geq \dE$,
which shows that the rows of $\E$ indeed generate an $[n,\ell,\dE]$ code.
As a result, any row of $\E$, as well as $\E$ itself, has $\Fq$-rank $\dE$.
\end{IEEEproof}

Due to the rank-metric Singleton bound, the minimum distance of the error code is upper bounded by $\dE \leq \tpub-\ell+1$. The work factor of \cite{tillich2018attack} is greater than RSD of Loidreau's system if $\dE > \tfrac{d-1}{2\lambda}$.
To gain in security level (or to reduce the key size), we must choose a suitable $[\tpub,\ell,\dE]$ code with
\begin{equation}
\tfrac{d-1}{2\lambda} < \dE \leq \tpub-\ell+1 \ . \label{eq:dE_condition}
\end{equation}
An obvious choice would be a Gabidulin code attaining the upper bound.
However, we will show in Appendix~\ref{app:Gabidulin_codes_not_suitable} that in this case, the error code $\rspace{\E}$ can be distinguished from a random code, which might be a weakness.

In the next section, we will show that it suffices to choose a random code as the error code since its minimum distance attains the upper bound in \eqref{eq:dE_condition} with high probability, cf.~\cite{neri2018MRD}.

As an alternative, one can use structured codes that arise from codes whose minimum distance is close to the upper bound. However, such codes have not been studied in the literature and, hence, this paper provides a motivation to study these codes. We will formally state the research problem in the conclusion.

\section{Using Random Error Codes}
In this section, we show that by choosing $\AE$ uniformly at random among all full-rank matrices in $\Fqm^{\ell\times\tpub}$, one obtains an $[n,\ell,\tpub-\ell+1]$ error code with high probability. For this choice of $\AE$, we then analyze the decryption failure probability.

\subsection{Probability of $\E$ Generating an $[n,\ell,\tpub-\ell+1]$ Code}
\begin{theorem}[Probabilities for MRD codes~\cite{neri2018MRD}]
  Let $\X \in \Fqm^{k\times(n-k)}$ be randomly chosen. Then
  \begin{equation*}
    \Pr\big[ \rspace{ \begin{mymatrix} \I_{k} | \X \end{mymatrix} } \text{is an MRD code} \big] \geq 1 - kq^{kn-m},
  \end{equation*}
  where $\I_{k}$ denotes the $k\times k$ identity matrix.
  \label{thm:mrd_prob}
\end{theorem}
Note that for practical parameters, it might not be feasible to determine the minimum rank distance of the chosen code since the fastest-known algorithms to compute the minimum rank distance are exponential in the code parameters.
\begin{proposition}
  Let $\E = \AE \BE$, where $\AE$  is drawn uniformly at random among all full-rank matrices in $\Fqm^{\ell \times \tpub}$ and $\BE$ uniformly at random among all full-rank matrices in $\Fq^{\tpub \times n}$. Then the probability that $\E$ is a generator matrix of a $[n,\ell,\tpub-\ell+1]$ code is $ \geq 1 - \ell q^{\ell\tpub-m}$.
\label{prop:E_MRD}
  \end{proposition}
  \begin{IEEEproof}
It follows directly from Theorems~\ref{thm:E_construction} and~\ref{thm:mrd_prob}.
  \end{IEEEproof}
  Note that if the inverse of the probability that $\AE$ is not MRD, i.e., $\ell^{-1}q^{m-\ell\tpub}$, is above the security level, this choice of the error does not decrease the security of the system. We take this into account for the choice of the proposed parameters and show the values in Table~\ref{tab:security_para}.
  
  \subsection{Decryption Failure Probability}
  The decryption algorithm fails if and only if the decoding of the interleaved Gabidulin code fails.
  \begin{lemma}\label{lemma:dim_prod}
    Let $\mathcal{B}$ be a fixed subspace and $\mathcal{A}$ a subspace generated by $\alpha$ random and linearly independent elements of $\Fqm$. Then,
    \begin{equation*}
      \Pr[\dim(\mathcal{A} \times \mathcal{B})= \alpha \beta] \geq 1 - \alpha q^{-(m-\alpha\beta)}. 
      \end{equation*}
    \end{lemma}
    \begin{IEEEproof}
See~\cite[Proposition 3.3]{gaborit2019LRPC}.
      \end{IEEEproof}

\begin{theorem}\label{thm:decode}
 Let $\Etilde = \AE \Btilde$, where $\AE$ is chosen as random full-rank matrix of $\Fqm^{\ell\times\tpub}$ and $\B$ as a random matrix of $\Fq^{\tpub \times n}$. Further let $\dim(\langle \AE_{i,1},\hdots,\AE_{i,\tpub}\rangle_{\Fq} \times \mathcal{V}) = \lambda \tpub$ for $i = 1,\dots,\ell$. Then correcting $\Etilde\P$ in $\IntGabcode{\ell;n,k}$ succeeds with probability
  \begin{equation*}
\geq \sum_{\tprime=\ell}^{\tpub\lambda} \frac{(1-\frac{4}{q^{m}})\big(1-\frac{q^{m\ell}}{q^{m\tprime}}\big)^{\ell}}{q^{\lambda\tpub n}} \prod_{i=0}^{\tprime -1} \frac{(q^{\tpub\lambda}-q^{i})(q^{n}-q^{i})}{q^{\tprime} -q^{i}}.
    \end{equation*}
\end{theorem}
\begin{IEEEproof}
The error that has to be decoded during decryption can be written as $\Etilde\P = \Aprime \Bprime$, where the $i$-th row of $\Aprime \in \Fqm^{\ell\times \lambda \tpub}$ is a basis of the product space $\langle \AE_{i,1},\hdots,\AE_{i,\tpub}\rangle_{\Fq} \times \mathcal{V}$ and $\Bprime \in\Fq^{\lambda\tpub \times n}$. Since $\dim(\langle \A_{i,1},\hdots,\A_{i,\tpub}\rangle_{\Fq} \times \mathcal{V}) = \lambda \tpub$ and $\Btilde$ is random, the matrix $\Bprime$ can be seen as random element of $\Fq^{\lambda\tpub\times n}$ and $(\E\P)_{i,j}$ as random element of $\langle \A_{i,1},\hdots,\A_{i,\tpub}\rangle_{\Fq} \times \mathcal{V}$, see~\cite[Proposition 4.3]{gaborit2019LRPC}. Thus, when applying the interleaved decoder proposed in~\cite{Loidreau_Overbeck_Interleaved_2006,Overbeck_Diss_InterleveadGab}, the probability of correcting $\Etilde\P$ successfully is
\begin{equation*}
\geq \sum_{\tprime=\ell}^{\tpub\lambda} (1-4q^{-m})\big(1-q^{-m(\tprime-\ell)}\big)^{\ell} \Pr [ \rank_{\Fq}(\Etilde\P) = \tprime ].
\end{equation*}
Further since $\rank_{q}(\Aprime) = \lambda \tpub$, the probability $\Pr[ \rank_{\Fq}(\Etilde\P) = \tprime ]$ is equal to the probability that the random matrix $\Bprime$ has rank $\tprime$~\cite[Proposition 4.3]{gaborit2019LRPC}, i.e.,
\begin{equation*}
\Pr [ \rank_{\Fq}(\Etilde\P) = \tprime ] = \frac{1}{q^{\lambda\tpub n}}\prod_{i=0}^{\tprime -1} \frac{(q^{\tpub\lambda}-q^{i})(q^{n}-q^{i})}{q^{\tprime} -q^{i}}.
\end{equation*}
\end{IEEEproof}
Note that the error in Theorem~\ref{thm:decode} is not necessary full-rank. However, it seems possible to adapt the proof of the bound in~\cite{Loidreau_Overbeck_Interleaved_2006,Overbeck_Diss_InterleveadGab} to random full-rank errors, where we conjecture that the lower bound on the success probability will be higher in case of full-rank errors. Based on this conjecture

the decryption algorithm in Section~\ref{sec:system} fails with probability
  \begin{align*}
    \leq 1- \sum_{\tprime=\ell}^{\tpub\lambda}  &\frac{(1-\frac{4}{q^m})\big(1-\frac{q^{m\ell}}{q^{m\tprime}}\big)  (1 - \tpub \frac{q^{\lambda\tpub}}{q^m})}{q^{\lambda\tpub n}}\\
    &\prod_{i=0}^{\tprime -1} \frac{(q^{\tpub\lambda}-q^{i})(q^{n}-q^{i})}{q^{\tprime} -q^{i}}.
    \end{align*}

We believe that the latter bound on the decryption failure is not tight since 1) $\dim(\langle \A_{i,1},\hdots,\A_{i,\tpub}\rangle_{\Fq} \times \mathcal{V}) = \lambda \tpub$ is not a necessary condition to successfully decode but only required for the correctness of Theorem~\ref{thm:decode} and 2) the bound was derived for $\E\P$ that might not have full rank. Nevertheless, for the parameters proposed in Table~\ref{tab:security_para}, even the inverse of this loose upper bound on the decryption failure rate is below the claimed security levels.

\section{Potential Key Size Reduction}
For the error construction proposed in Proposition~\ref{prop:E_MRD}, we propose parameters for (post-quantum) levels of security of $80$, $128$, $196$ and $256$ bit with respect to the known attacks in Table~\ref{tab:security_para}. The explicit work factors, the inverse of the probability that $\AE$ is not MRD denoted by $\text{WF}_{\AE}:=\log_{2}\ell^{-1}q^{m-\ell\tpub}$, the rate $k/n$, the key size $q^{k(n-k)}$ and the upper bound on the decryption failure $\Pfail$ in bits are presented for  $\dE = \tpub -\ell +1$.

\section{Conclusion}
  In this paper, we proposed a rank-metric McEliece-type cryptosystem based on applying the interleaving approach of Elleuch \emph{et al.} on Loidreau's cryptosystem. We analyzed possible attacks and showed that structural attacks are as hard as for Loidreau's system but an additional decoding attack is facilitated by interleaving. The efficiency of the latter attack can be reduced by choosing the error matrix as a generator matrix of a code with large minimum distance. We suggested design rules of the system and proved that depending on the parameters, a random construction of the error matrix fulfills the requirements with high probability. For this choice of the error, we derived upper bounds on the decryption failure and presented valid parameter sets that permit to decrease the key sizes.

\subsection*{Related Open Research Problem}

Note that \eqref{eq:dE_condition} does not restrict the code generated by $\AE$ to be MRD but also allows codes whose minimum distances are close to $\tpub-\ell+1$.
Since only little is known about non-MRD codes, the cryptosystem proposed here gives motivation to an interesting new research direction:

\begin{openproblem}\label{openprob:near-MRD_codes}
Given an extension field $\Fqm$, $n \leq m$, $k <n$, and $d = n-k+1-\varepsilon$, for some $\varepsilon \in \mathbb{N}$, $\varepsilon \ll n-k$,
find a rank-metric code with parameters $[n,k,d]$ over $\Fqm$ with efficient decoder, which---vaguely stated---cannot be distinguished from a random rank-metric code as easily as a Gabidulin code (cf.~Appendix) below).
\end{openproblem}

\appendix

\section{A Distinguisher for Errors from Gabidulin Codes}\label{app:Gabidulin_codes_not_suitable}

In this section, we show that choosing $\AE$ (cf.~Theorem~\ref{thm:E_construction}) to be a generator matrix of a Gabidulin code, results in an error code (i.e., the code spanned by the rows of $\E$) that is distinguishable from a random error matrix.
Although this does not directly lead to an explicit attack, which e.g., recovers the error matrix, this might be a weakness of ciphers obtained from these $\AE$.

We use the fact that the augmented matrix obtained by vertically concatenating $\Gpub$ and the cipher $\cipher$, has the same row space as the same construction with $\Gpub$ and the unknown error $\E$, i.e.,
$\rspace{\GaugY } = \rspace{\GaugE }$.
Thus, the augmented matrix might reveal the structure of the error matrix $\E$ by applying the following operator to it, as we will see in the following.

\begin{definition}[$q$-Sum]
Let $\Code[n,k]$ be a linear code over $\Fqm$ and $i \in \mathbb{N}_0$. Then, the ($i^{\mathrm{th}}$) $q$-sum of $\Code$ is defined by
\begin{equation*}
\LambdaOp_i(\Code) = \Code + \Code\qpow{1} + \dots + \Code\qpow{i}.
\end{equation*}
\end{definition}

\subsection{Distinguishing the Augmented Code}
\label{ssec:distinguisher_C}

We first state the following lemma.

\begin{lemma}\label{lemma:q-sum_C-ext}
Let $\E$ be constructed as in Theorem~\ref{thm:E_construction}, where $\AE$ is a generator matrix of a Gabidulin code.
Then,
\begin{equation*}
\dim(\LambdaOp_i(\Caug))  \leq \min\{(i+1)k + \min\{\ell+i,\tpub\}, n\} .
\end{equation*}
\end{lemma}

\begin{IEEEproof}
  By definition
$\LambdaOp_i(\Caug) = \LambdaOp_i(\rspace{\Gpub}) + \LambdaOp_i(\CodeE)$. 
Since $\AE$ is a generator matrix of a $[\tpub,k]$ Gabidulin code,
$\dim(\LambdaOp_i(\CodeE)) = \min\{\ell+i,\tpub\}$.
  Thus,
  {\small
  \begin{align*}
    \dim(\LambdaOp_i(\Caug)) &= \min\{\dim(\LambdaOp_i(\rspace{\Gpub})) + \min\{\ell+i,\tpub\}, n\} \\
    &\leq \min\{(i+1)k + \min\{\ell+i,\tpub\}, n\}. \qquad \qquad \IEEEQEDhere
  \end{align*}
}
\end{IEEEproof}

If $\AE$ in Theorem~\ref{thm:E_construction} is chosen to be a random full-rank matrix, we have
$\dim \LambdaOp_i (\rspace{\E}) = \min \{(i+1) \ell,\tpub \}$
with high probability. Hence, by the same arguments as in Lemma~\ref{lemma:q-sum_C-ext}, the overall augmented code has dimension
\begin{align*}
  \dim(\LambdaOp_i(\Caug)) =& \min \{\dim(\LambdaOp_i(\rspace{\Gpub})) \\
                                 &+  \min \{(i+1) \ell,\tpub\}, n \} \\
\leq& \min \{(i+1)k+\min \{(i+1) \ell,\tpub\},n\}.
\end{align*}
By Lemma~\ref{lemma:q-sum_C-ext}, for $2k+\min\{\ell+1,\tpub \}<n$ (which simply means that for $i>0$, $\min \{(i+1)k+\min\{\ell+i,\tpub\},n\}<n$), the dimension of $\LambdaOp_i(\Caug)$ with a Gabidulin code matrix $\AE$ is smaller than the respective dimension when using a random $\AE$, with high probability.
Hence, it can be distinguished.

\subsection{Distinguishing the Dual Augmented Code}
We study the dual of the augmented matrix.
\begin{lemma}
Let
\begin{equation*}
  \Caug^{\bot} := \rspace{
  \begin{mymatrix}
    \Gpub \\
    \cipher
    \end{mymatrix} }^{\bot}= \rspace{
  \begin{mymatrix}
    \Gpub \\
    \E
    \end{mymatrix} }^{\bot}, 
  \end{equation*}
  then $ \Caug^{\bot} = \rspace{\Gpub}^{\bot} \cap \rspace{\E}^{\bot}= \rspace{\Hpub} \cap \rspace{\HE}$.
\end{lemma}
\begin{IEEEproof}
 For the code $\Caug^{\bot}$ it holds that
  \begin{align*}
    \Caug^{\bot} &= \big \{ \c' : \c' \begin{mymatrix} \Gpub^{\top} \E^{\top} \end{mymatrix} = \ve{0}_{k+\ell} \big\} \\
                                   &= \{ \c' : \c' \Gpub^{\top} = \ve{0}_k \} \cap  \{\c': \c'  \E^{\top} = \ve{0}_\ell \} \\
                                   &= \rspace{\Gpub}^{\bot} \cap \rspace{\E}^{\bot},
  \end{align*}
  where $\vec{0}_i$ denotes the all-zero vector of length $i$.
\end{IEEEproof}

\begin{lemma}\label{lem:HE_form}
There is an $\HE$ of the form
\begin{equation*}
\HE = \begin{mymatrix}
\AE^\bot \Binv \\
\Bker
\end{mymatrix} \in \Fqm^{n-\ell \times n} \ ,
\end{equation*}
where $\Binv \in \Fq^{\tpub \times n}$ has $\Fqm$-rank $\tpub$, $\Bker \in \Fq^{(n-\tpub) \times n}$ has $\Fqm$-rank $n-\tpub$, $\AE^\bot \in \Fqm^{(\tpub-\ell) \times \tpub}$ and $(\AE^\bot \Binv^\top \BE^\top)$ is a parity-check matrix to $\AE$.
\end{lemma}

\begin{IEEEproof}
Since $\BE \in \Fq^{\tpub \times n}$ is of full rank and defined over $\Fq$, we can find a basis $\Bker \in \Fq^{(n-\tpub) \times n}$ of its right kernel. Note that $\Bker$ has full $\Fq$- and $\Fqm$-rank.
By the basis extension theorem, we can extend the linearly independent rows of $\Bker$ into a full basis of $\Fq^n$.
These further $\tpub$ basis element form the rows of $\Binv$.
Note that also $\Binv$ has full $\Fq$- and $\Fqm$-rank and any non-zero vector in the row space of $\Binv$ is linearly independent to the rows of $\Bker$.
Hence, also the rows of $\AE^\bot \Binv$ are linearly independent of the rows of $\Bker$, which, together with the fact that $\AE^\bot$ has full rank, shows that $\HE$ has full $\Fqm$-rank $n-\ell$.

It remains to show that the rows of $\HE$ are in the right kernel of $\E$.
The rows of $\Bker$ fulfill this because $\E = \AE \BE$ and $\Bker$ is a basis of the right kernel of $\BE$.
For the first $\tpub-\ell$ rows of $\HE$, we check:
\begin{align*}
\E (\AE^\bot \Binv)^\top &= \AE \BE \Binv^\top (\AE^\bot)^\top = \ve{0} \ ,
\end{align*}
which is true since $\AE^\bot \Binv^\top \BE^\top$ is a parity-check matrix with respect to $\AE$.
\end{IEEEproof}

\begin{remark}
Note that $\AE^\bot$ as in Lemma~\ref{lem:HE_form} is a generator matrix of a $[\tpub,\tpub-\ell]$ Gabidulin code since $\AE^\bot \Binv^\top \BE^\top$ is one (the dual code of a $[\tpub,\ell]$ Gabidulin code is a $[\tpub,\tpub-\ell]$ Gabidulin code, cf.~\cite{Gabidulin_TheoryOfCodes_1985}) and $\Binv^\top \BE^\top$ is an invertible matrix over $\Fq$ (which means that we just need to use different evaluation points in the Gabidulin code).
\end{remark}

\begin{lemma}\label{lemma:HE}
Let $\E$ be defined as in Theorem~\ref{thm:E_construction}, then
\begin{equation*}
  \dim\LambdaOp_i\big(\rspace{\HE}\big) \leq \min \{n-\ell+i,n\}.
\end{equation*}
\end{lemma}

\begin{IEEEproof}
We use Lemma~\ref{lem:HE_form}. Since $\Binv$ and $\Bker$ are over $\Fq$, $\Binv^q = \Binv$ and $\Bker^q = \Bker$. Further, $\dim \Lambda_i(\AE^\bot) = \min\{\tpub-\ell+i,\tpub\}$, cf.~\cite{Overbeck_Diss_InterleveadGab}, since $\AE^\bot$ is a generator matrix of a $[\tpub,\tpub-\ell]$ Gabidulin code. Thus, $  \dim\LambdaOp_i\big(\rspace{\HE}\big) \leq \min \{n-\ell+i,n\}$.
\end{IEEEproof}

\begin{lemma}
Let $\Hpub$ be a parity-check matrix of an $[n,k]$ code generated by $\Gpub$. Then,
$  \dim\LambdaOp_i\big(\rspace{\Hpub}\big) = \min \{(i+1)(n-k),n\},$
with high probability.
\label{lemma:Hpub}
\end{lemma}

\begin{lemma}
Let $\Caug^{\bot} =  \rspace{\Hpub} \cap \rspace{\HE}$. Then,
\begin{equation*}\label{lemma:Cdual}
  \Caug^{\bot} + \hdots +   \big( \Caug^{\bot}\big)^{[i]}  \subseteq  \rspace {
\begin{mymatrix}
  \Hpub \\
  \vdots \\
    \Hpub^{[i]}
    \end{mymatrix} }
    \cap
    \rspace {
    \begin{mymatrix}
  \HE \\
  \vdots \\
    \HE^{[i]}
    \end{mymatrix} }.
\end{equation*}
\end{lemma}

\begin{IEEEproof}
We have $\Caug^{\bot} \subseteq \rspace{\Hpub} $, thus $\Caug^{\bot} + \hdots +   \big(\Caug^{\bot}\big)^{[i]}  \subseteq \rspace{[\Hpub^\top, {\Hpub^{[1]}}^\top, \dots, {\HE^{[i]}}^\top]^\top}$.
The same holds for $\rspace{\HE}$, which proves the claim.
\end{IEEEproof}

\begin{theorem}
  Let $\Caug^{\bot} = \rspace{ \begin{mymatrix} \Gpub^{\top} \cipher^{\top}\end{mymatrix}^{\top}}^{\bot}$.
    Then,
$      \dim(\LambdaOp_i(\Caug^{\bot})) \leq \min\{n-l+i,(i+1)(n-k),n\}$.
  \end{theorem}
  \begin{IEEEproof}
The proof follows directly by Lemmas~\ref{lemma:HE}, \ref{lemma:Hpub} and \ref{lemma:Cdual}.
    \end{IEEEproof}

In summary, by choosing $\AE$ to be a generator matrix of a Gabidulin code, the error code $\E$ can be distinguished from an error matrix with random $\AE$.
This does not imply an explicit attack on the system, but indicates that there might be a weakness in this case.
The distinguisher must also be considered when constructing codes from Open Research Problem~\ref{openprob:near-MRD_codes}.

\bibliographystyle{IEEEtran}
\bibliography{main}

\end{document}